\documentclass[aps,prd,twocolumn,superscriptaddress,amsfont,amssymb,amsmath,nofootinbib,showpacs,balancelastpage]{revtex4-1}
\pdfoutput=1
	\usepackage[english]{babel}
	\usepackage{amsmath,amssymb,amsxtra}					
	
	\usepackage{graphicx}
\usepackage[latin1]{inputenc}
\usepackage{amsmath}
\usepackage{amsfonts}
\usepackage{appendix}
\usepackage{amssymb}
\usepackage{epsfig}
\usepackage{graphicx}
\usepackage{dcolumn}
\usepackage{bm}
\usepackage{epstopdf}
\usepackage{color} 
\usepackage[usenames,dvipsnames,svgnames]{xcolor}
\usepackage[colorlinks=true,
            linkcolor=red,
            urlcolor=gray,
            citecolor=blue]{hyperref}
\usepackage{hyperref}
\usepackage{multirow}
\usepackage{float}
\usepackage{hyperref}

\usepackage{subfigure}

\usepackage{enumitem}

\def\3nab{\tilde{\nabla}}

\def\be {\begin{equation}}
\def\ee {\end{equation}}
\def\ba {\begin{align}}
\def\ea {\end{align}}

\def\bc {\begin{center}}
\def\ec {\end{center}}
\def\case#1/#2{\frac{#1}{#2}}

\newcommand{\bver}{\begin{verbatim}}
\newcommand{\ever}{\end{verbatim}}

\newcommand{\bea}{\begin{eqnarray}}
\newcommand{\eea}{\end{eqnarray}}
\newcommand{\beaa}{\begin{eqnarray*}}
\newcommand{\eeaa}{\end{eqnarray*}}

\def\case#1/#2{\textstyle\frac{#1}{#2}}

\newcommand{\lsim}   {\mathrel{\mathop{\kern 0pt \rlap
  {\raise.2ex\hbox{$<$}}}
  \lower.9ex\hbox{\kern-.190em $\sim$}}}
\newcommand{\gsim}   {\mathrel{\mathop{\kern 0pt \rlap
  {\raise.2ex\hbox{$>$}}}
  \lower.9ex\hbox{\kern-.190em $\sim$}}}
\newcommand{\Od}{{\cal O}}
\begin{document}
\title{Isotropic charged cosmologies in infrared-modified electrodynamics}

\author{Jorge F.\ Soriano}
\email{jfernandezsoriano@gradcenter.cuny.edu}
\affiliation{Department of Physics, Graduate Center, City University of New York, New York, NY 10016, USA and Department of Physics and Astronomy, Lehman College, City University of New York, Bronx, NY 10468, USA}
\author{Antonio L.\ Maroto}
\email{maroto@ucm.es}
\affiliation{Departamento de F\'{\i}sica Te\'orica and Instituto de F\'isica de 
Part\'iculas y del Cosmos, Universidad Complutense de Madrid, 28040 
Madrid, Spain}

\pacs{98.80.-k}



\begin{abstract} 
It has long been known that the covariant formulation of quantum electrodynamics
conflicts with the local description of states in the charged sector. Some of the solutions to this problem amount to modifications of the subsidiary conditions below 
some arbitrarily low photon frequency. Such infrared modified theories have been shown to lead to  Maxwell equations modified with an additional 
 classical electromagnetic current induced by the quantum charges. The induced
 current only has support for very small frequencies and  cancels the effects of the physical charges on large scales. In this work we explore the 
 possibility that this de-electrification effect could allow  for the existence
 of isotropic charged cosmologies, thus evading the stringent limits 
 on the electric charge asymmetry of the universe. We consider a simple model 
 of infrared-modified scalar electrodynamics in the cosmological context and find that  the 
  charged sector generates a new contribution to the energy-momentum tensor whose
  dominant contribution at late times is a cosmological constant-like term. 
  If the charge asymmetry 
 was generated during inflation, the limits on the asymmetry parameter
in this model in order not to produce a too-large cosmological constant 
 are very stringent $\eta_Q <10^{-131}- 10^{-144}$ for a number of e-folds
 $N=50-60$ in typical models.  However if the charge imbalance is produced 
 after inflation, the limits are relaxed in such a way that 
 $\eta_Q<10^{-43}(100 \,\mbox{GeV}/T_Q)$, with $T_Q$ the temperature at which the asymmetry was generated. If the charge asymmetry has ever existed and the associated electromagnetic fields  vanish in the asymptotic future, the limit can be further reduced to $\eta_Q<10^{-28}$.
\end{abstract} 


\maketitle
\section{Introduction}

One of the long-standing questions in cosmology, dating back to the works of Bondi and
Lyttleton in the late fifties \cite{Lyttleton}, is the possibility that 
the universe could have a net electric charge density. It soon became apparent   
that this kind of charged cosmologies, even respecting large-scale
homogeneity, are  necessarily 
anisotropic.  Indeed, 
it is well-established in the context of Maxwell electrodynamics that  the presence of a non-vanishing
charge density $\rho$ generates electric fields
such that ${\bf \nabla} \cdot {\bf E}=\rho$. Even if we assume that both, the electric field and the  charge density, are spatially uniform within our Hubble horizon, the corresponding
electromagnetic energy-momentum tensor is nonetheless anisotropic. 
This implies a 
departure from the Robertson-Walker geometry which can conflict with 
the observed isotropy of different backgrounds. Thus, introducing the charge asymmetry parameter as $\eta_Q=n_Q/n_\gamma$, where the charge density is $\vert e \vert n_Q$, it has been shown that the isotropy of the cosmic microwave background imposes a limit $\eta_Q \,\lsim\, 10^{-30}$,  whereas the isotropy of the observed cosmic ray
distribution sets $\eta_Q \,\lsim \, 10^{-39}$ \cite{Orito}. On the other hand,
the electromagnetic interaction  with the 
electrostatic potential generated by the net charge density induces an
effective mass shift for any charged particle present in the cosmic plasma. 
These shifts introduce 
changes in the nucleosynthesis mechanism which can be translated 
into very stringent limits on the charge asymmetry $\eta_Q \,\lsim \,10^{-43}$ \cite{Masso}. 
The mentioned constrains were obtained ignoring the high conductivity 
of the cosmic plasma. When conductivity effects are taken into account
the limits can be improved, setting $\eta_Q\,\lsim \,10^{-35}$ for the CMB case  \cite{Ferreira}. 

On the theory side, several models have been proposed to generate a cosmological  charge imbalance. In \cite{gauge1,gauge2,gauge3}  spontaneous breaking of gauge invariance was considered,  the symmetry being restored at late times in order to comply with present 
experimental results. Other mechanisms involve the generation of charge by the presence of a photon mass \cite{Dolgov}
or prior to the GUT phase transition in brane-world scenarios. Quantum fluctuations 
of charged fields during inflation have also been considered in \cite{scalar1,scalar2,scalar3}
to  generate charge fluctuations on super-Hubble scales.

Even though the possibility of having a charge asymmetry compatible 
with current observations seems to be very limited in the context
of Maxwell electrodynamics, this is not the 
case in modified electromagnetic theories. Thus, 
Barnes \cite{Barnes} realized for the first time that the Proca generalization of electrodynamics, which propagates and additional longitudinal polarization 
for the photon,  admits the possibility that the universe could
possess a net electric charge density uniformly distributed throughout space, while possessing no electric or magnetic fields, thus allowing for homogeneous and 
isotropic Robertson-Walker solutions and evading most of the 
aforementioned limits. In particular, for constant photon mass and assuming charge conservation,  the electromagnetic energy-momentum tensor in this theory behaves, for homogeneous fields, as that of a perfect fluid with 
equation of state $p_P=\rho_P$ (see \cite{Zeldovich}) so that the energy density in a charged-dominated universe would scale as $a^{-6}$.  However, this scaling suggests that in order to avoid large 
contribution in the early universe, the potential contribution 
of charge in the present universe should be tiny. 
Extensions of these ideas in which the photon mass can depend on time were
considered in \cite{Woodard1,Woodard2} where a model for cosmic acceleration was proposed. 

In this work we re-examine the cosmology of a charged universe in Maxwell 
electrodynamics (with two propagating physical modes) from a different perspective. The well-known local and covariant formulation of quantum electrodynamics contains two fundamental ingredients: on one 
hand the dynamics, which is provided by Maxwell equations and, on the 
other, the constraints, which allow to eliminate the unphysical degrees of freedom, and  are given in the canonical formalism by the 
Gupta-Bleuler \cite{GB1,GB2} subsidiary conditions.
\begin{eqnarray}
(\nabla_\mu A^\mu)^{(-)}\vert \Phi\rangle=0
\end{eqnarray}
where $(\nabla_\mu A^\mu)^{(-)}$ is the negative frequency part of the 
 operator $\nabla_\mu A^\mu$ and $\vert \Phi\rangle$ denotes
 a physical state. 
However, 
this formulation of quantum electrodynamics
present certain difficulties in the charged sector which are known since the seventies \cite{Kulish, Ferrari}.
In particular Maison and Zwanziger  \cite{MZ} proved  a general result that states that there is no localized
charged state in covariant QED which satisfies the above subsidiary conditions.
In other words, either we abandon locality in the description of the charged
states or if we insist in a local description of charges we must assume
that all charged particles are produced from the decay of neutral states.

A possible way out of this limitation of covariant QED is the modification 
of the subsidiary condition in the infrared. An explicit implementation of these ideas have been presented by Zwanziger in \cite{Z} (see also
\cite{Haller,Z2}) 
and amounts to the introduction of an additional classical conserved current 
which is generated by the quantum current. In momentum space, this current has only support in the infrared, i.e. below the cutoff frequency, and cancels 
the effects of the quantum charges on very large scales. This property suggests that the cosmology of charged universes could exhibit important differences
in this kind of modified electrodynamics, and, in particular this opens the possibility of having isotropic charged solutions without including additional 
polarizations for the photon field. 

Even though Zwanziger model \cite{Z} is relatively old, its cosmological implications had not been analyzed so far. The aim of the present work  is precisely to evaluate the cosmological viability of that model in the context of charged cosmologies and dark energy models. We will find that the use of the modified formalism  generates new terms in the electromagnetic energy-momentum tensor which are not present in standard QED and whose dominant contribution at late-times is a cosmological constant term. By imposing such terms  to be compatible with current observations we
will set upper limits on the charge asymmetry of the universe in this scenario  

The paper is organized as follows: in Section II we review the infrared problems of covariant QED and their implications in the definition of charged states. In Section III, we present the Zwanziger subsidiary conditions and obtain the consistency condition for the classical current. In Section IV, 
we derive the equations of motion and energy-momentum tensors for the different components. Section V is devoted to the energy density and pressure of the scalar field. In Section VI, we calculate the induced electromagnetic energy density when different boundary 
conditions are imposed on the classical $b$ field and obtain the limits on the charge 
asymmetry. Finally in Section VII we present the main conclusions of the work.

\section{The infrared problem of perturbative QED}

Let us start by reviewing the long-standing infrared problem of QED and
its connection with the definition of charged particles. 

In the standard description of scattering processes in perturbative field theory, the assumption is made that in the initial and final asymptotic regions
the interactions can be switched off so that (in Minkowski space-time)  the fields can be expanded in plane waves solutions with the corresponding creation and annihilation free operators. In the interaction picture, these free field 
solutions can then be used to construct the corresponding interacting solution using the Green function for the interaction term. Even though this method can be straightforwardly applied in some toy models, in 
the case of unbroken gauge theories such as QCD or QED it exhibits important difficulties. Thus, in the QCD case, low-energy confinement
prevents the definition of asymptotically free quark states. 
In the case of QED the difficulty is less obvious since electrons are not confined, however  Fadeev and Kulish \cite{Kulish} showed that the 
masslessness of the photon implies that the electromagnetic interaction
does not decay sufficiently fast at long distances so as to neglect
it in the asymptotic regions. This residual interaction, implies that 
asymptotic charged fields can no longer be described as plane waves 
but they appear "dressed" by an electromagnetic field.  This in practice prevents the definition of local charged states in covariant QED. 

Let us then review in detail how the problem arises in the standard manifestly covariant formulation of QED in the Lorentz gauge. Here we will
closely follow the analysis in \cite{MZ}.  

The equations of motion in Minkowski space-time read \cite{Itzykson}:
\begin{equation} 
 \partial_\mu F^{\mu\nu}-\partial^\nu(\partial_\mu A^\mu)=J^\nu 
 \label{Max}
 \end{equation} 
where $J^\nu$ is the conserved current. In order to recover the classical Maxwell equation, the Lorentz condition $\partial_\mu A^\mu=0$
should be imposed. As is well known \cite{Itzykson}, this cannot be done at the operator
level but only in the weak sense given by the Gupta-Bleuler subsidiary 
conditions which in fact defines the physical Fock space of 
the theory:
\begin{eqnarray}
(\partial_\mu A^\mu)^{(-)}\vert \Phi\rangle=0
\label{GBM}
\end{eqnarray}
where $(\partial_\mu A^\mu)^{(-)}$ is the negative frequency part of the 
 operator $\partial_\mu A^\mu$ and $\vert \Phi\rangle$ denotes
 a physical state.

Rewriting equations (\ref{Max}) as
  \begin{equation} 
 \Box A_\mu=J_\mu 
 \end{equation} 
 and decomposing the external current as $J_\mu(x)=J_{\mu + }(x)+J_{\mu - }(x)$ with $J_{\mu \pm }=\theta(\pm \,x^0)J_\mu(x)$, the general 
 interacting solution $A_\mu(x)$ can be written in terms of the
 free solutions $A_\mu^f(x)$ satisfying $\Box A_\mu^f=0$ as
\begin{eqnarray}
A_\mu(x)&=&A_\mu^f(x)+\int \Delta^{ret}(x-y)\,J_{\mu +}(y)\,d^4y\nonumber \\
&+&
\int \Delta^{adv}(x-y)\,J_{\mu -}(y)\,d^4y\label{sol}
\end{eqnarray} 
where the retarded and advance propagators are
\begin{eqnarray}
\Delta^{ret}(x)=\frac{1}{(2\pi)^2}\delta(x^2)\delta(x^0)\nonumber\\
\Delta^{adv}(x)=\frac{1}{(2\pi)^2}\delta(x^2)\delta(-x^0)
\end{eqnarray} 
and the free field can be expanded in plane-wave solutions as
\begin{eqnarray}
A_\mu^f(x)=\frac{1}{(2\pi)^3} \int\frac{d^3k}{2\omega}\left(a_\mu(\vec k) e^{-i k x}+a_\mu^\dagger(\vec k) e^{i k x}\right)
\end{eqnarray} 
where $\omega=\vert\vec k\vert $, and $a_\mu$,  $a_\mu^\dagger$ denote the {\it free} annihilation and
creation operators satisfying
\begin{eqnarray}
[a_\mu(\vec k),a_\nu^\dagger(\vec k')]=-\eta_{\mu\nu}\,2\omega\, \delta^3(\vec k-\vec k')
\end{eqnarray} 
From (\ref{sol}) we can write
\begin{eqnarray}
\partial_\mu A^\mu(x)&=&\partial_\mu A^\mu_f (x)+\int \Delta(x-y)\,J^{0 }(y)\delta(y^0)\,d^4y \label{Lor}
\end{eqnarray} 
with 
\begin{eqnarray}
\Delta(x-y)&=&\Delta^{ret}(x-y)-\Delta^{adv}(x-y)\nonumber \\
&=&\frac{i}{(2\pi)^3} \int\frac{d^3k}{2\omega}\left(e^{-i k x}+ e^{i k x}\right)
\end{eqnarray}
Thus, substituting back (\ref{Lor}) in (\ref{GBM}), we get the Gupta-Bleuler subsidiary conditions in Fourier space
\begin{eqnarray}
(\omega\,(a_0(\vec k) - a_\parallel(\vec k))-\rho(\vec k))\vert \Phi\rangle=0 \label{GBF}
\end{eqnarray}
where $a_0(\vec k)$ and $a_\parallel(\vec k)=\frac{k^i}{\omega} a_i(\vec k)$ denote the temporal and longitudinal
annihilation operators of the {\it free} fields respectively and $\rho(\vec k)$
is the charge density operator in Fourier space, i.e
\begin{equation}
\rho(\vec k)=\frac{1}{(2\pi)^{3/2}}\int  e^{-i\vec k\vec x}\,J^0(0,\vec x)\,d^3x
\end{equation}
If $\rho(\vec k)$ is a smooth function, then 
\begin{equation}
\rho(0)=\frac{1}{(2\pi)^{3/2}}\int J^0(0,\vec x)\,d^3x=\frac{Q}{(2\pi)^{3/2}}
\end{equation}
with $Q$ the total charge. It is now possible to obtain solutions
of the subsidiary condition (\ref{GBF}) for the physical states $\vert \Phi \rangle$ in the form
\begin{eqnarray}
\vert \Phi \rangle=\exp{\left(-\frac{1}{2}\int(a_0^\dagger(\vec k) + a_\parallel^\dagger(\vec k))\frac{\rho(\vec k)}{\omega}\frac{d^3k}{2\omega}\right)}\vert \Psi \rangle
\end{eqnarray}
with 
\begin{eqnarray}
\vert \Psi \rangle= {\cal F}[a_0^\dagger(\vec k) - a_\parallel^\dagger(\vec k),a_\perp^\dagger(\vec k)]\vert 0\rangle
\end{eqnarray}
where $a_\perp^\dagger(\vec k)$ denotes the two transverse creation operators and ${\cal F}$ is an analytical function. Thus, the norm 
squared of a physical state is given by
\begin{eqnarray}
\langle \Phi\vert \Phi\rangle= \exp{\left(\int \frac{\vert \rho(\vec k)\vert^2}{\omega^2} \frac{d^3k}{2\omega} \right)}
\label{norm}
\end{eqnarray}
Notice that since $\rho(\vec k)$ is a smooth function with $\rho(0)=Q/(2\pi)^{3/2}$, the above integral is infrared divergent for $Q\neq 0$.
This means that the Gupta-Blueler condition has no Fock space solution with finite norm in the charged sector \cite{MZ}. 

One of the possible solutions to this problem is the modification 
of the subsidiary condition in the infrared. In particular it can  
be seen \cite{MZ} that a modified condition given by:
\begin{eqnarray}
\left(\omega(a_0(\vec k) - a_\parallel(\vec k)) -\rho(\vec k)+Q \,f_c(\omega)\right)\vert \Phi\rangle=0
\label{mGB0}
\end{eqnarray}
where $Q$ is the charge operator and $f_c(\omega)$ is a cut-off function such that
$f_c(0)=(2\pi)^{-3/2}$ and $f_c(\omega)=0$ for $\omega>\omega_0$, defines a non-empty Fock space
of physical states.  Indeed, the new term can be 
seen as a classical  current which screens the 
effect of the quantum charges on large scales thus allowing 
for finite norm states in (\ref{norm}) and in this way avoids the infrared problem. Notice that this expression implies that the subsidiary conditions are only modified  
below an arbitrarily low frequency $\omega_0$, so that 
standard QED is recovered on small scales. In next section we will describe in detail the implementation of this modified electrodynamics in the Zwanziger model \cite{Z}.

\section{Zwanziger subsidiary conditions}

Let us consider a simple renormalizable scalar electrodynamics theory minimally coupled to gravity. The Lagrangian density of this model can be written as \cite{Z} 
\begin{multline} 
\mathcal L=-\frac14 F_{\mu\nu}F^{\mu\nu}-\frac{\lambda}{2}(\nabla_\mu A^\mu)^2
\\+(D_\mu \varphi)^*(D^\mu\varphi)-V(|\varphi|),\label{eq1.1}
\end{multline}
where $D_\mu=\partial_\mu+iqA_\mu$, and $q$ is the $U(1)$ charge of the scalar field. The fundamental fields appearing here are $A_\mu$, $\varphi$ and $g_{\mu\nu}$, and the action is \begin{equation} S[A_\mu,\varphi,g_{\mu\nu}]=\int \sqrt{\mathfrak g}\,\mathcal L\, d^4x,\label{eq1.2}\end{equation} with $\mathfrak{g}\equiv |\det(g_{\mu\nu})|$.

The corresponding equations of motion  are \begin{equation} \nabla_\alpha F^{\alpha\mu}+\lambda\nabla^\mu(\nabla_\alpha A^\alpha)=J^\mu\label{eq2.1}\end{equation} and \begin{equation}D_\mu(\sqrt{\mathfrak g}\,D^\mu\varphi)+\frac{1}{2}\sqrt{\mathfrak g}\, \frac{V'(|\varphi|)}{|\varphi|}=0,\label{eq2.2}\end{equation} where \begin{equation} J^\mu=iq[\varphi^*(D^\mu\varphi)-\varphi(D^\mu\varphi)^*]\label{eq2.3}\end{equation} is the $U(1)$ conserved current, i.e. such that $\nabla_\mu J^\mu=0$. We will see later how current conservation can be derived from (\ref{eq2.2}), as one would expect. Notice that although the action in (\ref{eq2.1}) is invariant under the restricted gauge transformations,
$A_\mu\rightarrow A_\mu+\partial_\mu \Lambda$
with transformation parameter satisfying $\Box \Lambda=0$, the scalar
sector is fully gauge invariant, so that current conservation is preserved.

Taking the divergence of (\ref{eq2.1}) and taking into account the 
conservation of the current, we get
\begin{eqnarray}
\Box (\nabla_\alpha A^\alpha)=0
\end{eqnarray}
Following \cite{Z}, we define the physical states $\vert \Phi\rangle$
as those given by the following modification of the Gupta-Bleuler condition:
\begin{eqnarray}
(\nabla_\alpha A^\alpha)^{(-)}(x)\vert \Phi\rangle=b^{(-)}(x)\vert \Phi\rangle
\label{mgb}
\end{eqnarray}
where $b(x)$ is a real c-number solution of the wave equation 
\begin{eqnarray}
\Box b(x)=0
\end{eqnarray}
which can be separated into its positive and negative frequency parts as
$b(x)=b^{(+)}(x)+b^{(-)}(x)$ and that in Fourier space would generate the cut-off function $f_c(\omega)$ in (\ref{mGB0}).

Then it can be shown that if  $b(x)$ satisfies:
 \begin{eqnarray}
\int \sqrt{g_\Sigma} \,\partial_\mu b(x)\, d\Sigma^\mu=\frac{q}{\lambda}
\label{cc}
\end{eqnarray}
with $\Sigma$ a constant-time hypersurface,  $g_\Sigma$ the metric on $\Sigma$, $d\Sigma^\mu$ the future-oriented volume element on $\Sigma$ and $q$ the charge
of the state $\vert \Phi\rangle$ then $\langle \Phi\vert \Phi\rangle\geq 0$, i.e. 
states satisfying (\ref{mgb}) have non-negative norm and such subspace is  
invariant under the action of observables ${\cal O}$ i.e. $\langle {\cal O}\Phi\vert {\cal O}\Phi\rangle\geq 0$. Notice that unlike \cite{Z} we are working 
with an arbitrary $\lambda$.

These conditions imply that for  the expectation value we get 
\begin{eqnarray}
\langle \Phi\vert (\nabla_\alpha A^\alpha)(x)\vert \Phi\rangle=b(x)
\label{exp}
\end{eqnarray}
so that the classical Maxwell equations are modified with the introduction of an 
additional classical current and can be written as \cite{Z}
\begin{eqnarray}
\nabla_\alpha F^{\alpha\mu}=J^\mu-\lambda\nabla^\mu b
\label{mMaxwell}
\end{eqnarray}
Notice that even though the subsidiary condition are modified, the  
gauge invariance of the scalar sector preserves the Ward identities of the 
theory, so that we do not expect any uncompensated production of temporal
and longitudinal photons in the theory. As a matter of fact, as shown 
\cite{Z}, 
all the cross sections formulae of standard QED are recovered in the
modified formalism. 

\section{Charged cosmologies}

We will apply the previous formalism in cosmology for the description of 
a homogeneous and isotropic universe with a uniform charge density. 
With that purpose we consider a spatially-flat Robertson-Walker metric 
\begin{equation} ds^2=dt^2-a^2d\vec x^2,\label{eq2.4}
\end{equation} 
where $a\equiv a(t)$ is the scale factor and hence, $\sqrt{\mathfrak{g}}=a^3$. 

We would like to find non-trivial solutions in which the matter fields (scalar and electromagnetic fields) are also homogeneous and isotropic. With that purpose we will look for solutions 
of the classical equation of motion where the vector fields cannot point in 
any spatial direction, so that $A_i=0$. Then, the only remaining component of the field is $A_0$.  By homogeneity, spatial derivatives of any field must vanish, which gives us $\partial_i A_\mu=0$ and $\partial_i\varphi=0$. With these conditions, current conservation reads \begin{equation} a^3(t)J_0(t)=\kappa, \label{eq2.5}\end{equation} being $\kappa$
the constant comoving charge density of the scalar field. 

For the spatial hypersurface of constant $t$, we have $\sqrt{g_\Sigma}=a^3(t)$ and $d\Sigma^\mu=d^3 x (1,0,0,0)$, so that the consistency condition (\ref{cc}) reads
 \begin{eqnarray}
\int_V a^3 \,\partial_0 b(x)\, d^3 x=\frac{1}{\lambda} \int_V a^3 \,J_0\, d^3 x
\end{eqnarray}
with $V$ a given comoving volume. In the cosmological context it is natural to impose
that $b$ is a homogeneous field, i.e.
$b=b(t)$ which means that in Fourier space $b(k)$ only has contribution from the zero mode, i.e. the corresponding cutoff frequency would be essentially $k_0= 0$. In other words,
the modification of the subsidiary condition would only affect the zero mode electromagnetic fields. For the rest of states, the standard Gupta-Bleuler condition is recovered. Notice that since the tip of the light-cone
$\omega_0=0$ is Lorentz invariant, we do not expect any modification in 
the subsidiary condition in a boosted frame. If we further assume that the consistency condition
is valid for an arbitrary cosmological volume $V$, then we finally obtain
\begin{eqnarray}
\lambda\,\partial_0 b(t)=J_0
\label{cc0}
\end{eqnarray}
which implies that the right-hand side of (\ref{mMaxwell}) vanishes, i.e.  
even though we have a net charge density, the presence of the new current cancels 
its effects on cosmological scales. This means a vanishing Faraday tensor on large scales, so that it is possible to get exact homogeneous and isotropic Robertson-Walker solutions. This \emph{de-electrification} of the electric current, which is decoupled from the electromagnetic fields,  is different from the degravitation mechanism \cite{Dvali} of the cosmological constant from gravity. Although both cases resort to 
infrared modifications of the theory, in the gravitational case it is the dynamics 
rather than the subsidiary conditions what is modified  in order to absorb the vacuum energy contribution.

On the other hand, notice that introducing (\ref{eq2.5}) into the Maxwell equations (\ref{eq2.1}) and taking into account the isotropy and homogeneity of $A_\mu$, we can write a $\varphi$-independent equation of motion for the classical electromagnetic field: 
\begin{equation} \lambda\, \partial_0(\nabla_\mu A^\mu)=\kappa a^{-3}, \label{eq2.6.0}\end{equation} 
which is compatible with (\ref{cc0}) and can be rewritten as:
\begin{equation} \lambda\, (\ddot A_0 +3H\dot A_0+3\dot H A_0)=\kappa a^{-3}, \label{eq2.6}\end{equation} 
 being $H\equiv \dot a/a$ the Hubble parameter.

The scalar field can be written in terms of a modulus and a phase: $\varphi=f\,e^{i\theta}$. Introducing this expression in (\ref{eq2.2}) gives, after splitting the resulting equation in its real and imaginary parts, 
\begin{equation} 
\ddot f+3H\dot f-(\dot\theta+q A_0)^2f+\frac{1}{2}V'(f)=0\label{eq2.8}
\end{equation} 
and 
\begin{multline} 
\ddot\theta f+2\dot f\dot\theta+3H\dot\theta f+2qA_0\dot f\\+q(\dot A_0+3HA_0)f=0.  \label{eq2.9}
\end{multline}

In order to simplify this expression, we write the zero component of the current density in terms of $f$ and $\theta$ as 
\begin{equation} 
\kappa a^{-3}=J_0=-2q(\dot\theta+qA_0)f^2,\label{eq2.10}
\end{equation} 
in such a way that (\ref{eq2.8}) becomes  
\begin{equation} 
\ddot f+3H\dot f-\frac{\kappa^2}{4q^2a^6f^3}+\frac{1}{2}V'(f)=0.\label{eq2.11}
\end{equation} 
Now, we can write (\ref{eq2.9}) in a different way as 
\begin{equation} 
\frac{1}{f a^3}\frac{d}{dt}[a^3f^2(\dot\theta+qA_0)]=0.\label{eq2.12}
\end{equation} 
Thus, comparison with (\ref{eq2.10}) shows  us that this equation of motion is completely equivalent to current conservation, as was mentioned before.

The stress-energy tensor is obtained by the variation of the action with respect to the metric as 
\begin{equation} 
T^{\mu\nu}=-\frac{2}{\sqrt{\mathfrak{g}}}\frac{\delta S}{\delta g_{\mu\nu}}.\label{eq2.13}
\end{equation} 
We get the complete stress-energy tensor as a sum of the contributions from the scalar and  electromagnetic fields, $T^{\mu\nu}=T^{\mu\nu}_{\varphi}+T^{\mu\nu}_A$. For our model, we get its symmetrized components as 
\begin{multline} 
T^{\mu\nu}_\varphi=2(D^{(\mu}\varphi)^*(D^{\nu)}\varphi)\\-g^{\mu\nu}((D_\alpha\varphi)^*(D^\alpha\varphi)-V(|\varphi|)),\label{eq2.14}
\end{multline} 
and 
\begin{multline} 
T^{\mu\nu}_A=-F^{\mu\alpha}{F^\nu}_\alpha+2\lambda A^{(\mu}\nabla^{\nu)}(\nabla_\alpha A^\alpha)\\-g^{\mu\nu}\left[-\frac{1}{4}F_{\alpha\beta}F^{\alpha\beta}+\frac{\lambda}{2}(\nabla_\alpha A^\alpha)^2\right.\\\left.\vphantom{\frac{\lambda}{2}}+\lambda A^\alpha \nabla_\alpha(\nabla_\beta A^\beta)\right].\label{eq2.15}
\end{multline} 
Although gauge covariant derivatives appear in (\ref{eq2.14}), we see that in our case, using (\ref{eq2.10}), $T^{\mu\nu}_\varphi$ can be written in terms of  the modulus of the scalar field $f$ only. As a matter of fact, the only non-vanishing components of (\ref{eq2.14}) and (\ref{eq2.15}) are the energy densities and pressures, obtained as $\rho_{(\alpha)}={{T_{(\alpha)}}^0}_0$ and $p_{(\alpha)}=-{{T_{(\alpha)}}^i}_i$, where $(\alpha)$ stands for $\varphi$ or $A$. Thus  we get 
\begin{subequations}
\begin{equation} 
\rho_\varphi=\dot f^2+\frac{\kappa^2}{4q^2a^6f^2}+V(f),\label{eq2.16a}
\end{equation}
\begin{equation} 
p_\varphi=\dot f^2+\frac{\kappa^2}{4q^2a^6f^2}-V(f),\label{eq2.16b}
\end{equation}\label{eq2.16}
\end{subequations} 
and 
\begin{subequations}
\begin{equation}
\rho_A=\kappa A_0 a^{-3}-\frac{\lambda}{2}(\nabla_\mu A^\mu)^2\label{eq2.17a},
\end{equation}
\begin{equation}
p_A=\kappa A_0 a^{-3}+\frac{\lambda}{2}(\nabla_\mu A^\mu)^2\label{eq2.17b},
\end{equation}\label{eq2.17}
\end{subequations} 
where we have made use of (\ref{eq2.6}) to get the standard Coulomb interaction $\kappa A_0 a^{-3}$ term. Notice the difference with respect to the standard electromagnetic energy density in QED since, in addition to the Coulomb term, in  the Zwanziger model a new extra contribution is present which is proportional to $b^2$ according to (\ref{exp}).

\section{Energy and pressure of the scalar field}

In order to obtain the scaling behaviour of the energy density of the scalar 
field, we will particularize the scalar field potential to the simplest case corresponding to a mass term, \begin{equation}V(f)=\frac{1}{2}m^2f^2.\label{eq3.1}\end{equation} 

We will solve (\ref{eq2.11}) numerically, so we rewrite it here in a dimensionless form.  To do so, we define a dimensionless time as $\tau\equiv mt$ and  define $\tau$ valued fields and parameters as the barred ones: 
\begin{equation}
\begin{split}
\bar f(\tau)&\equiv\frac{q}{m}f(t),\\\bar a(\tau)&\equiv a(t),\\\bar H(\tau)&\equiv\frac{1}{\bar a(\tau)}\frac{d}{d\tau}\bar a(\tau).
\end{split}\label{eq3.3}
\end{equation} In the following, we shall omit the $\tau$ dependence, which is obvious in barred quantities.

\begin{figure}\centering\includegraphics[width=\linewidth]{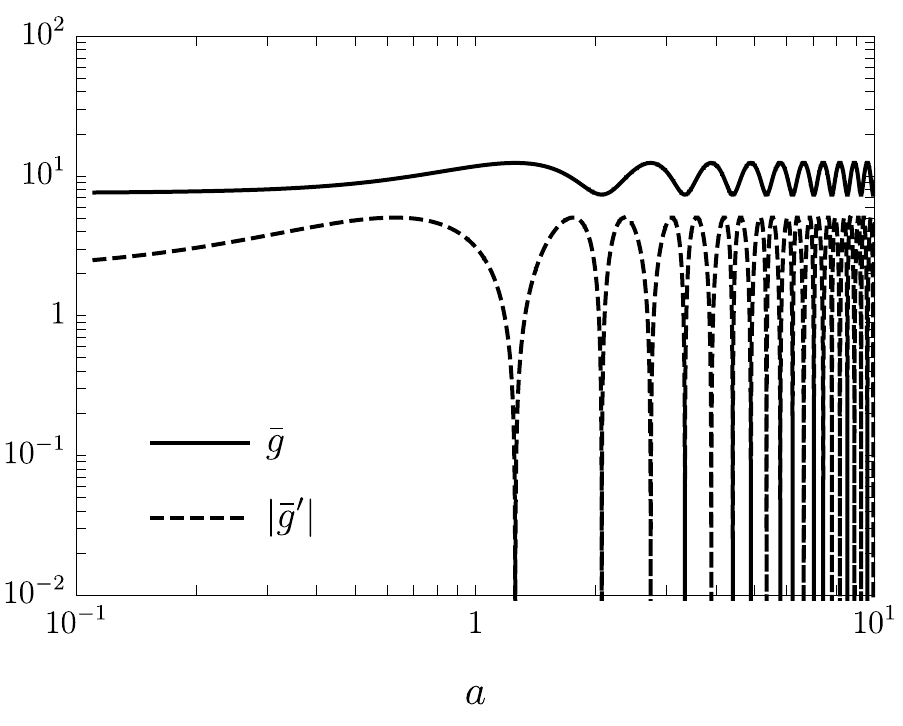}\caption{Evolution of the modified scalar field in terms of the scale factor.}\label{fig1}\end{figure}

Now, the equation of motion (\ref{eq2.11}) with the mass potential can be written as \begin{equation} 
\bar f''+3\bar H\bar f'-\frac{\bar \kappa^2}{\bar a^6\bar f^3}+\bar f=0,\label{eq3.4}\end{equation} 
where the prime means $\tau$ derivative, and we have defined the dimensionless constant \begin{equation} 
\bar \kappa\equiv \frac{\kappa q }{2m^3}.\label{eq3.5}
\end{equation} 
In order to simplify the numerical evaluation of (\ref{eq3.4}) we define 
\begin{equation} \bar g\equiv \bar a^{3/2}\,\bar f\label{eq3.6}
\end{equation} 
so that the equation of motion becomes 
\begin{equation} \bar g''-\frac{\bar \kappa^2}{\bar g^3}+\left[1-\frac{9\bar H^2}{4}-\frac{3\bar H'}{2}\right]\bar g=0,\label{eq3.7}
\end{equation} 
where the damping term $\bar g'$ does not  appear. 
We will solve for different cosmological eras assuming $\bar a\propto \tau^n$,
so that we get the final equation 
\begin{equation} 
\bar g''-\frac{\bar \kappa^2}{\bar g^3}+\left(1-\frac{3n(3n-2)}{4\tau^2}\right)\bar g=0\label{eq3.8}
\end{equation} 
Thus for radiation dominated era with $n=1/2$, we get
\begin{subequations}
\begin{equation}
\bar g''-\frac{\bar \kappa^2}{\bar g^3}+\left(1+\frac{3}{16\tau^2}\right)\bar g=0,\label{eq3.9}
\end{equation}
whereas for matter domination with $n=2/3$, it reads
\begin{equation}
\bar g''-\frac{\bar \kappa^2}{\bar g^3}+\bar g=0.\label{eq3.10}
\end{equation}
\end{subequations} 
Solving these equations numerically we get, for a wide range of initial conditions   an oscillatory behaviour for the field around a constant value at late times 
when $m\gg H$, as showed in Fig.\ref{fig1}, both for $\bar g$ and $\bar g'$. This translates into an oscillatory 
behaviour for the field $\bar f$ with an amplitude that decays with the scale factor as $a^{-3/2}$.

If we introduce this behaviour in the scalar field energy density (\ref{eq2.16a}), with the mass potential (\ref{eq3.1}), we will find that it decays 
at late time as \begin{equation} \rho_f\propto\frac{1}{a^3},\label{eq3.11}\end{equation} so that the scalar field has a matter like behaviour as expected \cite{Turner}. This means that 
provided that initial conditions ensure that it is subdominant with respect to the total matter density it
will be a subdominant component at all times.

\section{Energy and pressure of the electromagnetic field}
\subsection{Setting the background}
In order to study analytically the behaviour of both the energy density and pressure in (\ref{eq2.17}) we should solve the equation of motion for $A_0(a)$, given by (\ref{eq2.6}). Changing time derivatives to $a$-derivatives, and  integrating twice and once  respectively  from some 
initial value $a_i$ we can get, after some manipulations, \begin{subequations}\begin{multline}A_0(a)=\left(\frac{a_i}{a}\right)^3A_0|_i+\frac{\nabla_\mu A^\mu|_i}{H_0 a^3}(\mathcal F-\mathcal F_i)\\+\frac {\kappa}{\lambda H_0^2 a^3}[\mathcal G-\mathcal G_i-\mathcal I_i(\mathcal F-\mathcal F_i)],\label{eq4.1a}\end{multline} and \begin{equation}\nabla_\mu A^\mu=\left.\nabla_\mu A^\mu\right|_i+\frac {\kappa}{\lambda H_0}(\mathcal I-\mathcal I_i),\label{eq4.1b}\end{equation}\label{eq4.1}\end{subequations} where we have defined the primitive functions \begin{subequations}\begin{equation}\mathcal F\equiv \int\frac{a'^2}{E(a')}da'\label{eq4.2a}\end{equation}\begin{equation}\mathcal G\equiv \int\frac{a'^2}{E(a')}da'\int\frac{da''}{a''^4 E(a'')}\label{eq4.2b}\end{equation}\begin{equation}\mathcal I\equiv\int\frac{da'}{a'^4 E(a')},\label{eq4.2c}\end{equation}\label{eq4.2}\end{subequations} and $H(a)\equiv H_0 E(a)$. These primitives are evaluated in $a$ except if they have some subscript that indicates a particular constant scale factor.

\begin{figure}[h]\centering\includegraphics[width=.7\linewidth]{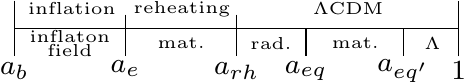}\caption{}\label{fig2}\end{figure}

In order to obtain explicit expressions for the integrals, we will consider the 
standard cosmological behaviour with an initial inflationary phase followed by a reheating
phase connecting with the radiation era of standard $\Lambda$CDM cosmology. 
We will assume for simplicity that the energy density scales as matter during the 
reheating phase \cite{KolbTurner}. This background scheme is depicted in Figure \ref{fig2}.

We consider a quasi- de Sitter inflationary phase with almost constant Hubble parameter $H_I$  which can be estimated from the Friedmann equation as \begin{equation} H_I=\sqrt\frac{V_I}{3M_p^2}\label{eq4.5}\end{equation}  where $M_p^2=1/(8\pi G)$ and $V_I^{1/4}$ is the scale of 
inflation. On the other hand we have, $a_b=e^{-\mathcal N}a_e$, where $\mathcal N$ is the
total number of inflation e-folds. In addition, since we are assumming the reheating era to be matter dominated, where $\rho\propto a^{-3}$, it is possible to write \begin{equation} a_e=\left(\frac{\rho(a_{RH})}{\rho(a_e)}\right)^\frac13a_{RH}.\label{eq4.7}\end{equation} Taking  $ \rho(a_e)=V_I$ and that $\rho(a_{RH})=\frac{\pi^2}{30}g_* T_{RH}^4$, one can approximate \begin{equation} a_e\simeq\left(\frac{T^4_{RH}}{V_I}\right)^\frac13 a_{RH},\label{eq4.8}\end{equation} where we have ignored numerical factors of order one.
Finally, we need $a_{RH}$. Assuming adiabatic expansion after reheating  we have  \begin{equation} a_{RH}=\frac{T_{eq}}{T_{RH}}a_{eq},\label{eq4.9}\end{equation} where $T_{eq}\simeq 0.83$ eV and $a_{eq}=2.8\times 10^{-4}$,  the standard values for the temperature and scale factor at matter-radiation 
equality \cite{KolbTurner}. Thus we see that the details of the inflationary and reheating phases are  encapsulated in the three parameters $T_{RH}$, $V_I$ and $\mathcal N$.

Thus, up to order-one numerical factors, the final function to integrate is 
\begin{equation} 
E(a)=\left\{\begin{array}{cc}
\displaystyle{\frac{1}{H_0}\sqrt{\frac{V_I}{M_p^2}},}&\displaystyle{a_b<a<a_e}\\[.5cm]
\displaystyle{\frac{T_{RH}^2}{H_0 M_p }\left(\frac{T_{eq}}{T_{RH}}\right)^\frac32 \left(\frac{a_{eq}}{a}\right)^\frac32,}&\displaystyle{a_e<a<a_{RH}}\\[.5cm]
\displaystyle{\sqrt{\Omega_\Lambda+\frac{\Omega_M}{a^3}+\frac{\Omega_R}{a^4}},}&\displaystyle{a_{RH}<a,}
\end{array}\right.\label{eq4.10}\end{equation}
with \begin{subequations}\begin{equation} a_b=e^{-\mathcal N}\left(\frac{T_{RH}^4}{V_I}\right)^\frac13\frac{T_{eq}}{T_{RH}}a_{eq},\label{eq4.11a}\end{equation}\begin{equation}a_e=\left(\frac{T_{RH}^4}{V_I}\right)^\frac13\frac{T_{eq}}{T_{RH}}a_{eq},\label{eq4.11b}\end{equation}\begin{equation}a_{RH}=\frac{T_{eq}}{T_{RH}}a_{eq}.\label{eq4.11c}\end{equation}\label{eq4.11}\end{subequations}  

\subsection{Evolution of the energy density}
Now, we can study how the energy density of the electromagnetic part evolves in the background described by (\ref{eq4.10}) and (\ref{eq4.11}). This would depend on (a) the initial time where the charge density appears, and (b) on the boundary conditions that we set for both $A_0$ and $\nabla_\mu A^\mu$.

If we introduce the fields into the energy density, grouping terms adequately, we get 
\begin{multline}
\rho_A=\left[\kappa a_i^3 A_0|_i-\frac{\kappa}{H_0}\left(\nabla_\mu A^\mu|_i-\frac{\kappa\mathcal I_i}{\lambda H_0}\right)\mathcal F_i-\frac{\kappa^2 \mathcal G_i}{\lambda H_0^2}\right]\frac{1}{a^6}\\+\frac{\kappa}{H_0}\left(\nabla_\mu A^\mu|_i-\frac{\kappa\mathcal I_i}{\lambda H_0}\right)\left(\frac{\mathcal F}{a^6}-\mathcal I\right)\\-\frac{\kappa^2}{\lambda H_0^2}\left(\frac{\mathcal I^2}{2}-\frac{\mathcal G}{a^6}\right)\\-\frac{\lambda}{2}\left(\nabla_\mu A^\mu|_i-\frac{\kappa\mathcal I_i}{\lambda H_0}\right)^2.
\label{eq*4.11}\end{multline}


Let us study the evolution of the various energy density terms in different situations.

\subsubsection{Instantaneous charge density generation}
Let us first consider the case in which charge and  electromagnetic fields 
vanish initially and at some $a=a_i$ a net charge density is generated, so that we 
take $A_0|_i=0$ and $\nabla_\mu A^\mu|_i=0$ in (\ref{eq*4.11}) and obtain

\begin{subequations}\begin{equation}\rho_A=\frac{\kappa^2}{\lambda H_0^2}\left(\frac{\mathcal G-\mathcal G_i-\mathcal I_i(\mathcal F-\mathcal F_i)}{a^6}-\frac12 (\mathcal I-\mathcal I_i)^2\right),\label{eq4.13a}\end{equation}\begin{equation}p_A=\frac{\kappa^2}{\lambda H_0^2}\left(\frac{\mathcal G-\mathcal G_i-\mathcal I_i(\mathcal F-\mathcal F_i)}{a^6}+\frac12 (\mathcal I-\mathcal I_i)^2\right).\label{eq4.13b}\end{equation}\label{eq4.13}\end{subequations}

The value of $a_i$ will be determined by the charge generation mechanism. Thus, as mentioned 
before  a charge density could be generated, for example, during inflation due to some fluctuation of the charged scalar field \cite{scalar1,scalar2,scalar3}, or at a phase transition \cite{gauge1,gauge2,gauge3}   well inside the radiation era.

\begin{figure}\centering\includegraphics[width=\linewidth]{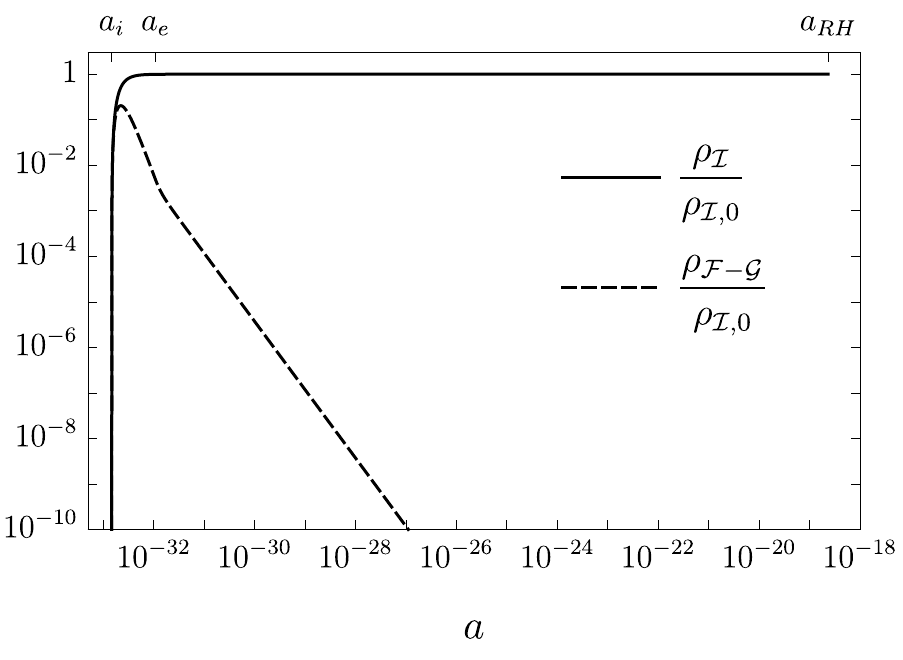}\caption{Behaviour of the terms in (\ref{eq*4.11}) normalized to the dominant value today, for a charge asymmetry generated during inflation era. Here $\rho_I \mathcal\propto(\mathcal I-\mathcal I_i)^2$ and $\rho_{\mathcal F -\mathcal G}\propto [\mathcal G-\mathcal G_i-\mathcal I_i(\mathcal F-\mathcal F_i)]a^{-6}]$.}\label{figgh}\end{figure}

Thus, for charge generation  occurring during  the inflation era ($a_i\ll a_e$), we can see that the terms appearing in (\ref{eq4.13}) behave as shown in Fig. \ref{figgh}. In this plot $\mathcal N_i\equiv\ln\frac{a_e}{a_i}=2$, i.e. only two inflation e-folds are needed to reach the
asymptotic behaviour. We can see that at the end of inflation, the terms containing $\mathcal I$ are  two orders of magnitude bigger than those with $\mathcal F$ and $\mathcal G$. After inflation, the difference remains increasing, and nowadays only the $\mathcal I$ terms are relevant. Moreover, we see that $\mathcal I-\mathcal I_i$ remains constant after that, i.e. $\mathcal I\simeq \mathcal I_e$ for $a>a_e$, so that the 
dominant contribution at late times is a cosmological constant component. 

Since in inflation one has the analytical expression \begin{equation}\mathcal I=-\frac{M_p H_0}{3\sqrt{V_I}}\frac{1}{a^3},\label{eq4.14}\end{equation} we can see that, at the end of inflation, \begin{equation} \mathcal I_e-\mathcal I_i=\frac{M_p H_0}{3\sqrt{V_I}}\frac{1}{a_i^3}\left(1-e^{-3\mathcal N_i}\right)\simeq -\mathcal I_i,\label{eq4.15}\end{equation} for big enough $\mathcal N_i$. Then, the energy density today is \begin{equation}\rho_{A,0}\simeq-\frac{\kappa^2}{2\lambda H_0^2}\mathcal I_i^2,\label{eq4.16}\end{equation} with \begin{equation}\mathcal I_i=-\frac{M_p H_0\sqrt{V_I}}{3T_{RH} T_{eq}^3}\frac{e^{3\mathcal N_i}}{a_{eq}^3},\label{eq4.17}\end{equation} where we have used (\ref{eq4.11b}) and $a_i=e^{-\mathcal N_i} a_e$.
Thus for $\lambda<0$, it is possible to have a positive  cosmological constant today.

We now want to study the range of values of the charge
density $\kappa$ which can provide a value for $\rho_A$ today consistent with dark energy density measurements. In order to do it, we use that \begin{equation}\rho_\Lambda=\Omega_\Lambda \rho_c=3 H_0^2 M_p^2\Omega_\Lambda.\label{eq4.18}\end{equation} 

Comparing this expression with (\ref{eq4.16}) and (\ref{eq4.17}), we get \begin{equation} -\frac{\kappa^2}{\lambda}\lsim\frac{54a_{eq}^6\Omega_\Lambda}{e^{6\mathcal N_i}}\frac{H_0^2 T_{RH}^2 T_{eq}^6 }{V_I}.\label{eq4.19}\end{equation}

Thus  we get for the constant comoving charge density
\begin{equation}\vert\kappa\vert\lsim\frac{\sqrt{54\Omega_\Lambda\vert \lambda\vert}}{z_{eq}^3 e^{3\mathcal N_i}}\frac{H_0T_{RH}T_{eq}^3}{\sqrt{V_I}}\label{eq4.20}\end{equation} 
so that
 \begin{equation}
 \vert\kappa\vert\lsim 10^{-78}e^{-3\mathcal N_i}\vert\lambda\vert^{1/2}
\left(\frac{T_{RH}}{10^6 \mbox{GeV}}\right)
\left(\frac{10^{16} \mbox{GeV}}{V_I^{1/4}}\right)^2
 \mathrm{eV}^3,\label{eq4.21}
 \end{equation} which allows to obtain, assuming individual particles of charge $e=\sqrt{4\pi\alpha}$, 
 the limit on the charge asymmetry  as   \begin{equation} \eta_Q\lsim[10^{-144},10^{-68}]\,
\vert\lambda\vert^{1/2}
\left(\frac{T_{RH}}{10^6 \mbox{GeV}}\right)
\left(\frac{10^{16} \mbox{GeV}}{V_I^{1/4}}\right)^2 \end{equation} for $\mathcal N_i\in[2,60]$.

\begin{figure}\centering\includegraphics[width=\linewidth]{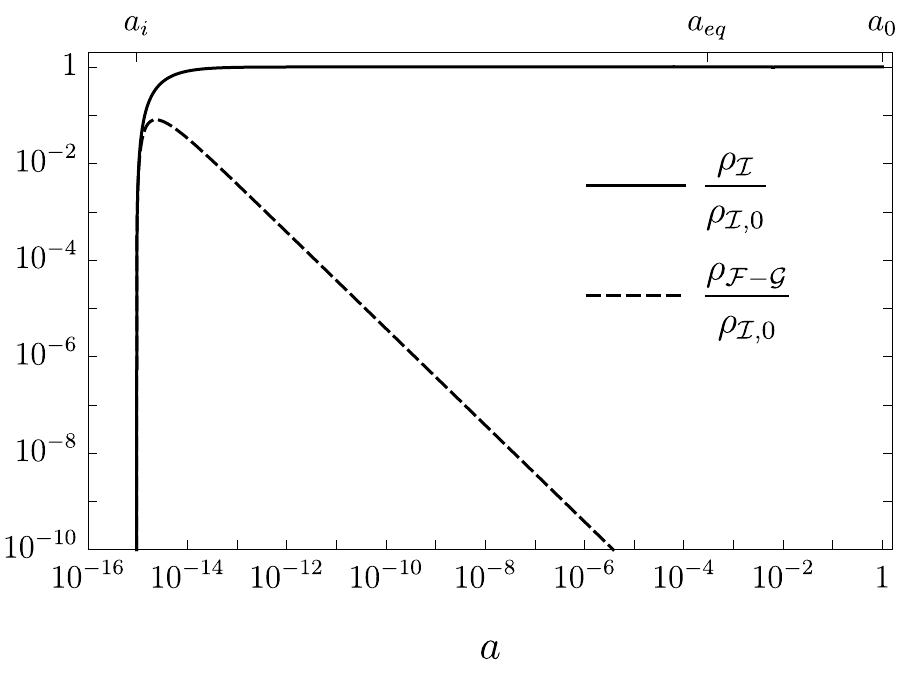}\caption{Same as in Fig. \ref{figgh} but for a charge asymmetry generated at $T_Q=T_{EW}=246$ GeV. }\label{fig.ew}\end{figure}

Let us now consider the case in which the 
charge generation takes place well inside the 
radiation era at a temperature $T=T_Q$, corresponding to an initial scale factor \begin{equation} a_i=a_{Q}=\frac{T_{eq}}{T_{Q}} a_{eq}.\label{eq4.23}
\end{equation} 
In Fig. \ref{fig.ew} we can see that, again, the $\mathcal I$ terms are several orders of magnitude bigger than the $\mathcal F$ and $\mathcal G$ ones. Moreover, after $a_{eq}$, $\mathcal I-\mathcal I_i$ has again a constant value until today, so that we can approximate the energy density today as the energy density in the radiation-matter equality. We use the analytical solution for $\mathcal I$ during a radiation dominated universe, 
\begin{equation} \mathcal I=-\frac{1}{\sqrt{\Omega_R} a},\label{eq4.24}
\end{equation} 
in order to write 
\begin{equation} \mathcal I_{eq}-\mathcal I_{i}=\frac{1}{\sqrt{\Omega_R}}\frac{1}{a_{eq}}\left(\frac{T_{Q}}{T_{eq}}-1\right)\approx-\mathcal I_{i}, \label{eq4.25}
\end{equation} 
and then, following the same steps  as in (\ref{eq4.18}-\ref{eq4.20}), one can estimate \begin{equation}\vert\kappa\vert=\sqrt{6\Omega_\Lambda \Omega_R\vert\lambda\vert}H_0^2 M_p a_{eq}\frac{T_{eq}}{T_{Q}}, \label{eq4.26} \end{equation} which,  gives the limit on the charge asymmetry  \begin{equation}\eta_Q \lsim 10^{-43}\vert\lambda\vert^{1/2}\left(\frac{100\;  \mbox{GeV}}{T_{Q}}\right).\label{eq4.27}\end{equation}
Thus we see that for $\vert \lambda\vert =\Od(1)$, a tiny charge
asymmetry of order $\eta_Q\simeq 10^{-43}$ produced at the electroweak phase transition would generate a cosmological constant compatible with observations. The possibility 
of generating a cosmological constant in the context of modified electrodynamics was considered in the uncharged case in \cite{BeltranMaroto1,BeltranMaroto2}. Notice that in the uncharged sector with $q=0$ i.e. $\kappa=0$, the only possible homogeneous solution of the consistency condition (\ref{cc}) is a constant $b$ field, which contributes to $\rho_A$ in (\ref{eq2.17a})  as
a pure cosmological constant.


\subsubsection{Vanishing fields in the asymptotic future}
Another type of solutions correspond to those in which charge density has ever been 
present but the induced 
electromagnetic field vanish asymptotically in the future as the charge 
density decrease as $\kappa\propto a^{-3}$, i.e.  
\begin{equation} 0=\lim_{a\to\infty} \nabla_\mu A^\mu=\nabla_\mu A^\mu|_i-\frac{\kappa\mathcal I_i}{\lambda H_0}+\frac{\kappa}{\lambda H_0}\lim_{a\to\infty}\mathcal I.\label{eq4.29}\end{equation} For $a\gg1$ only the $\Omega_\Lambda$ term will survive, and $\mathcal I$
 vanishes in the limit. Then this yields \begin{equation}\nabla_\mu A^\mu|_i=\frac{\kappa\mathcal I_i}{\lambda H_0}.\label{eq4.30}\end{equation} Now, taking the same limit in (\ref{eq4.1a}), one can see that it is automatically satisfied for any $A_0|_i$ taking the previous result for $\nabla_\mu A^\mu|_i$. By introducing (\ref{eq4.30}) in the energy density (\ref{eq*4.11}), we can see that this is equivalent to removing the constant mode, as one would expect. Moreover, the term $\mathcal F a^{-6}-\mathcal I$ is removed too. The energy density is then \begin{equation}\rho_A=\left[\kappa a_i^3 A_0|_i-\frac{\kappa^2 \mathcal G_i}{\lambda H_0^2}\right]\frac{1}{a^6}-\frac{\kappa^2}{\lambda H_0^2}\left(\frac{\mathcal I^2}{2}-\frac{\mathcal G}{a^6}\right),\label{}\end{equation} and the pressure  \begin{equation}p_A=\left[\kappa a_i^3 A_0|_i-\frac{\kappa^2 \mathcal G_i}{\lambda H_0^2}\right]\frac{1}{a^6}+\frac{\kappa^2}{\lambda H_0^2}\left(\frac{\mathcal I^2}{2}+\frac{\mathcal G}{a^6}\right).\label{}\end{equation} 

As mentioned above, we can safely neglect the term with $A_0|_i$ which makes the analysis easier. Now, let us define the dimensionless functions\begin{subequations}\begin{equation}\bar\rho_A\equiv \frac{\mathcal I^2}{2}-\frac{\mathcal G-\mathcal G_i}{a^6},\end{equation}\begin{equation}\bar p_A\equiv-\frac{\mathcal I^2}{2}-\frac{\mathcal G-\mathcal G_i}{a^6}.\end{equation}\end{subequations} We solve (\ref{eq4.2b}) and (\ref{eq4.2c}) for the different epochs 
\begin{enumerate}
\item Inflation:\begin{subequations}
\begin{align}
\bar\rho_A&=\frac{M_p^2H_0^2}{V_I}\left(\ln a+\frac{1}{6}+\mathcal G_i\right)\frac{1}{a^6},\label{eq4.34a}\\
\bar p_A&=\frac{M_p^2H_0^2}{V_I}\left(\ln a-\frac{1}{6}+\mathcal G_i\right)\frac{1}{a^6}.\label{eq4.34b}
\end{align}
\label{eq4.34}
\end{subequations}
\item Reheating:\begin{subequations}\begin{align}
\bar\rho_A&=\frac{4H_0^2M_p^2}{3T_{RH}T_{eq}^3}\frac{1}{(aa_{eq})^3}+\frac{\mathcal G_i}{a^6},\label{eq4.35a}\\
\bar p_A&=\frac{\mathcal G_i}{a^6}.\label{eq4.35b}\end{align}\label{eq4.35}\end{subequations}
\item Radiation\begin{subequations}
\begin{align}\bar\rho_A&=\frac{3}{4\Omega_R a^2}+\frac{\mathcal G_i}{a^6},\label{eq4.36a}\\
\bar p_A&=-\frac{1}{4\Omega_R a^2}+\frac{\mathcal G_i}{a^6}.\label{eq4.36b}\end{align}\label{eq4.36}\end{subequations}
\item Matter: 
\begin{subequations}
\begin{align}\bar\rho_A&=\frac{4}{9\Omega_M a^3}+\frac{\mathcal G_i}{a^6},\label{eq4.37a}\\
\bar p_A&=\frac{\mathcal G_i}{a^6}.\label{eq4.37b}\end{align}\label{eq4.37}\end{subequations}
\item Dark energy:\begin{subequations}
\begin{align}\bar\rho_A&=\frac{1}{3\Omega_\Lambda}\left(\ln a+\frac16+\mathcal G_i\right)\frac{1}{a^6},\label{eq4.38a}\\
\bar p_A&=\frac{1}{3\Omega_\Lambda}\left(\ln a-\frac16+\mathcal G_i\right)\frac{1}{a^6}.\label{eq4.38b}\end{align}\label{eq4.38}\end{subequations}
\end{enumerate}

\begin{figure}\centering\includegraphics[width=\linewidth]{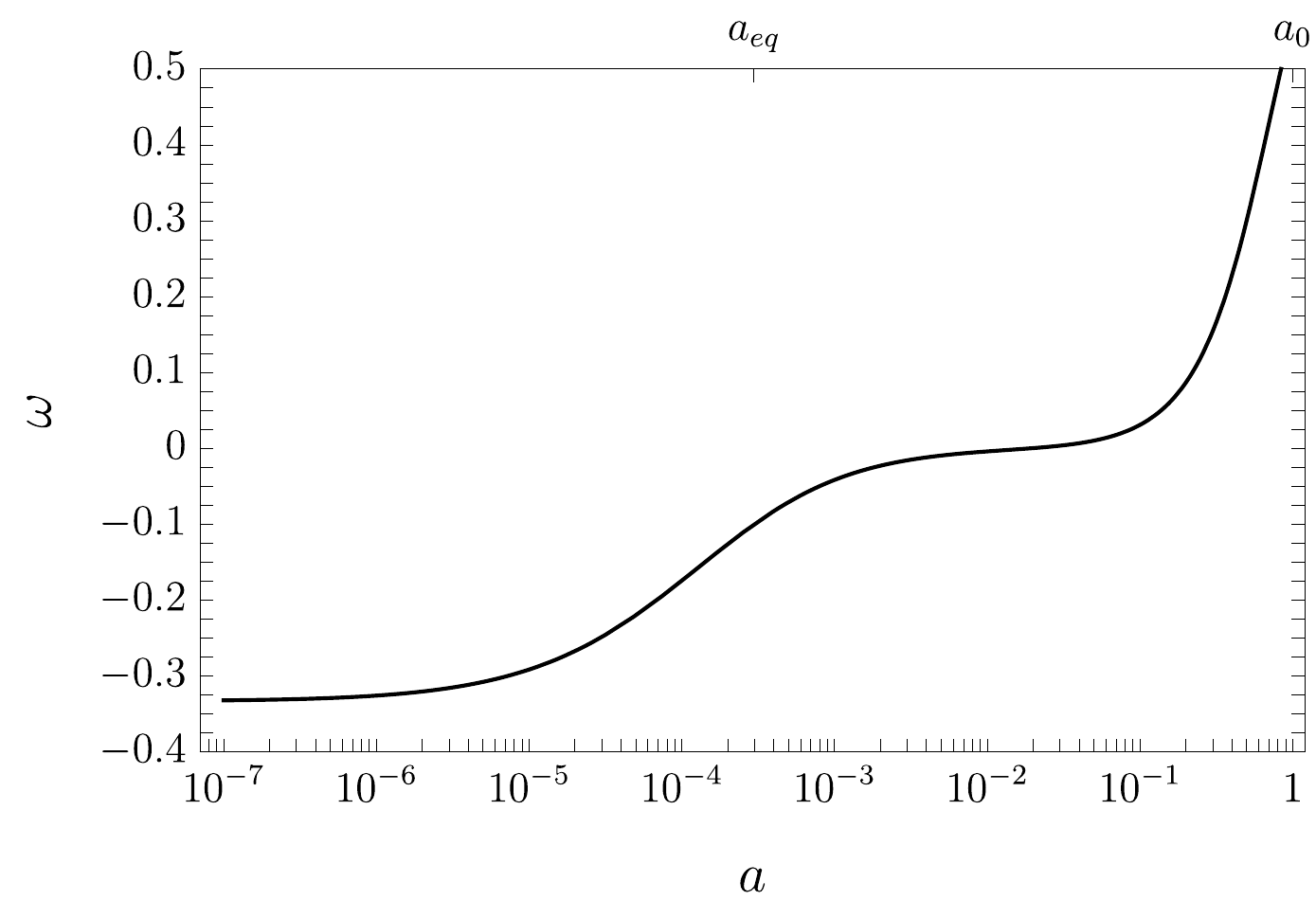}\caption{Evolution of the effective electromagnetic equation of state $w_A$.}\label{omegainf}\end{figure}

The evolution of the  equation of state is shown in Fig. \ref{omegainf}. We can see that 
far from the transition regions i.e. neglecting $\mathcal G_i$,  $w_A=-1/3$ in the radiation era,  $w_A=0$ in the matter era and
during the accelerated expansion eras, the behaviour goes as: \begin{equation}w_A(a)=1-\frac{2}{1+6\ln a},\label{eq4.39}\end{equation} 
that tends asymptotically to that of a stiff fluid $w_A=1$. 


\begin{figure}\centering\includegraphics[width=\linewidth]{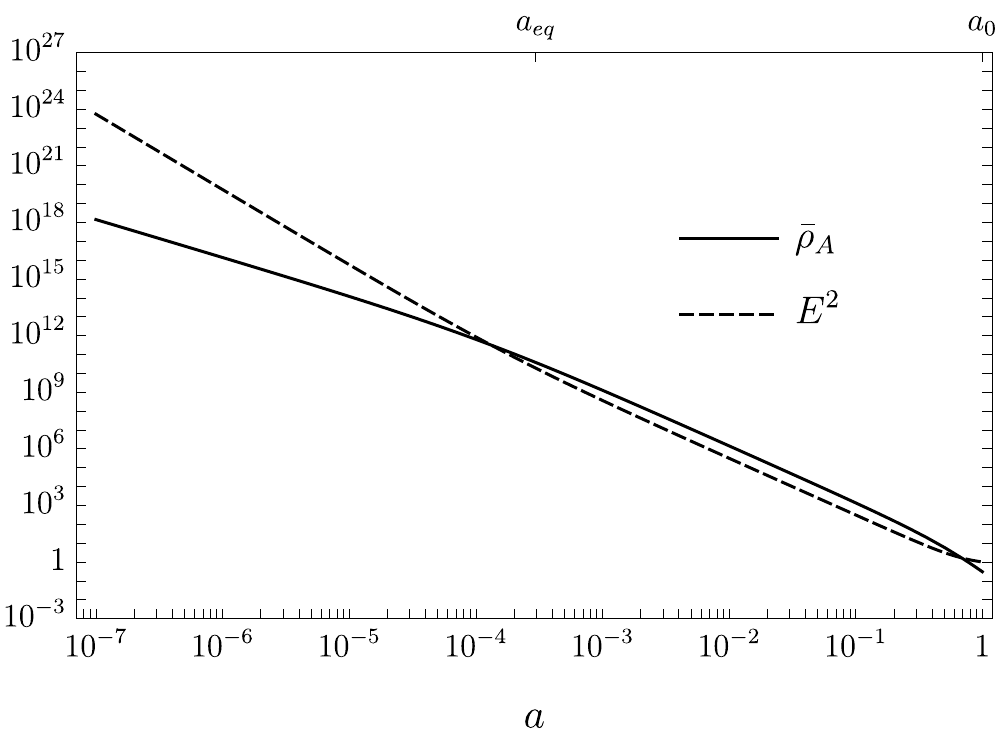}\caption{Evolution of the normalized electromagnetic energy density $\bar \rho_A(a)$  (solid) and normalized $\Lambda$CDM energy density $E^2(a)$ (dashed).}\label{dens}\end{figure}

In Fig. \ref{dens} we show the evolution of the dimensionless $\bar \rho_A$ compared to
 $E^2(a)$ which follows the scaling of $\rho_{\Lambda CDM}$. As we can see the maximum contribution occurs in the matter dominated era, when the ratio $R=\bar \rho/E^2$
reaches a maximum value $R_{max}=4.8$. Thus, in order for the charge-induced energy density not to spoil the predictions of standard $\Lambda$CDM, we impose $\rho_A\lsim 10^{-2} \rho_{\Lambda CDM}$ which implies
\begin{eqnarray}
R_{max}\frac{8\pi G}{3}\frac{\kappa^2}{\vert\lambda\vert H_0^4}\lsim 10^{-2}
\end{eqnarray}
which, in turn, can be translated into a limit on the charge asymmetry as
\begin{eqnarray}
\eta_Q\lsim 10^{-28} \vert\lambda\vert^{1/2}
\end{eqnarray}
This limit relaxes in several order of magnitude, the present bounds on the charge asymmetry
in standard Maxwell electrodynamics mentioned before.

\section{Conclusions}
We have explored the possibility of constructing homogeneous and isotropic cosmologies
with a non-vanishing charge density in the context  of modified Maxwell electrodynamics.
Unlike previous works which considered theories that include a small photon mass and thus propagate three degrees of freedom, we have limited ourselves to the Zwanziger model of electrodynamics, with two propagating polarizations, but 
with modified subsidiary conditions. The modification  affects only the physical 
photon Fock space in the infrared. 

We show that in the context of this model, the induced classical current counterbalance
the effects of the physical (quantum) charges so that the Faraday tensor vanishes
on cosmological scales, thus allowing for the construction of exact Robertson-Walker
geometries. Depending on the boundary conditions imposed on the classical $b(t)$ field, 
different scenarios are possible. Thus, if $b(t)$ vanishes at some initial time 
when the charge density is generated, then the dominant contribution to the 
electromagnetic energy-momentum tensor is a cosmological constant-like term. Imposing
the value of the induced constant to be smaller than the observed one 
sets stringent limits on the comoving charge density, which translates into
limits on the charge asymmetry  which can range from $\eta_Q \lsim 10^{-131}$ if charges are produced during inflation (for typical inflationary models) to 
$\eta_Q \lsim 10^{-43}\vert\lambda\vert^{1/2}(100\;  \mbox{GeV}/T_{Q})$ if the charge density is generated in the radiation era at a temperature $T_Q$. In the case in which the 
$b$ field vanishes asymptotically in the future when the charge density also vanishes, 
the cosmological constant-like term is absent and the dominant contribution 
appears as an extra matter density in the matter dominated era. Imposing again compatibility 
with the observed matter density sets a weaker limit $\eta_Q\lsim 10^{-28} \vert\lambda\vert^{1/2}$, several orders of magnitude below the limits in standard Maxwell electrodynamics. 

The de-electrification mechanism discussed in this work only takes place on cosmological scales, 
and,  a priori, could not prevent the appearance of  effects on smaller scales. However, 
the small charge densities suggest that such effects could be actually suppressed. Thus for example, in the highest density case, corresponding to   $\eta_Q\simeq 10^{-28}$, the corresponding density of charged particles today would be $n_Q\simeq 10^{-26}$ cm$^{-3}$.
Of course density perturbations would induce also charge density perturbations
which could be enhanced in high-density objects. The study of the evolution of these
charge fluctuations  is however beyond the scope of the present work.

\section*{acknowledgments}

This work has been partially supported by MINECO grant FIS2016-78859-P(AEI/FEDER, UE) and by  Red Consolider MultiDark FPA2017-90566-REDC. The research of J.F.S. is supported by U.S. National Science Foundation (Grant PHY-1620661).

\end{document}